\documentclass[aps,prl,preprint,groupedaddress]{revtex4}

\usepackage[dvips]{graphicx} 

\usepackage{dcolumn}

\usepackage{color}

\begin{document}



\title{Neutron reactions in accreting neutron stars: A new pathway to efficient crust heating}

\author{Sanjib S. Gupta}
\author{Toshihiko Kawano}
\author{Peter M\"{o}ller}
\affiliation{Theoretical Division, Los Alamos National Laboratories, Los Alamos, NM 87545}


\date{\today}

\begin{abstract}

In our calculation of neutron star crust heating we include several key new
model features. In earlier work electron capture (EC) only allowed neutron
emission from the daughter ground-state; here we calculate, in a deformed
QRPA model, EC decay rates to all states in the daughter that are allowed by
Gamow-Teller selection rules and energetics.  The subsequent branching ratios between the
1n,\dots,xn channels and the competing $\gamma$-decay are calculated in a Hauser-Feshbach
model. Since EC accesses excited states, many more neutrons are emitted in our
calculation than in previous work, leading to accelerated reaction flows. In our multi-component plasma 
model a
single (EC,xn) reaction step can produce several neutron-deficient nuclei, each
of which can further decay by (EC,xn). Hence, the neutron emission occurs more continuously with increasing depth
as compared to that in a 
one-component plasma model.

\end{abstract}

\pacs{26.60.Gj, 95.85.Nv, 97.60.Jd, 97.80.Jp}

\maketitle

Nuclear reactions in the crust of a neutron star (NS) form an important bridge between the extensively studied nuclear burning on the surface and the less well understood neutrino cooling processes in the core, where exotic states of matter can exist. Observation of accreting neutron stars in binary systems is proving to be a promising new tool to probe the thermal evolution of the inner core and its neutrino cooling rate. In particular, the ignition of ``X-ray superbursts" (XRSB) due to $^{12}$C fusion has been shown to be sensitive to the thermal profile of the neutron star crust\cite{Brown04}. Similarly, neutron stars exhibiting intermittent periods of rapid accretion (``transients'') have a quiescent luminosity which is thought to be powered by the heat released in the crust \cite{BBR98}. 

Surface burning on an accreting NS by the ``rp-process" (rapid proton capture) in  a Type-I X-ray Burst (XRB) \cite{WH04, schatz} produces heavy elements near the proton drip-line. As these ashes are buried by subsequent accretion successive electron capture (EC) reactions occur as the electron chemical potential $E_{\mathrm{F}}$ increases rapidly with depth. In pioneering work a decade ago, Haensel and Zdunik outlined these EC reactions in a one-component plasma (OCP) and found that the heat released in EC was not substantial -- most of the energy release occurred much deeper in the crust through pycnonuclear fusion reactions \cite{HZ90,HZ03}. 
More recently in a multi-component plasma (MCP) simulation, Gupta \emph{et.\frenchspacing al.} found that EC to excited states and subsequent $\gamma$-deexcitation increased the heating at shallow depths \cite{Gupta}. In this letter we report that EC captures to highly excited daughter states also lead to neutron emission processes near the neutron drip-line that are vastly accelerated when the level schemes of deformed neutron-rich nuclei are incorporated as opposed to when emissions are restricted to ground states (g.s.). These reactions fundamentally change the direction and speed of the reaction pathway and alter the crustal heating profile.

For $E_{\mathrm{F}}\gtrsim27$ MeV there will be a neutron chemical potential $\mu_{\mathrm{n}}\gtrsim0$ and consequently a free neutron abundance. The depth $E_{\mathrm{F}}\simeq27$ MeV is called the ``neutron drip'' (ND) point.
Close to ND ($E_{\mathrm{F}}\sim25$ MeV), EC can occur into excited states with $E_{\mathrm{exc}} > S_{\mathrm{xn}}$ where $S_{\mathrm{xn}}$ is the separation energy for $x$ neutrons. These reactions are henceforth referred to as (EC,xn) reactions. 
Once (EC,xn) is allowed at $E_{\mathrm{F}}\sim25$ MeV, any EC is always accompanied by some neutron emission since the $x\lesssim3$ channels are always open and the $x=0$ channel (``pure EC") is highly suppressed. 
Post-ND the products of (EC,xn) can be highly EC-unstable (``superthreshold") at $(Z-1)$ since the neutron number $N$ has also been substantially lowered. Although some of these nuclei will recapture neutrons, yet the favored $N$ after (n,$\gamma)$ will \emph{not} be as high as the pure EC pathway would have permitted, due to the powerful non-thermal driving force removing neutrons. This force is $(E_{\mathrm{F}} - E_{\mathrm{thresh,g.s.}}-S_{\mathrm{xn}})$ and since it is much greater than $kT$ the superthreshold (EC,xn) rates are also independent of temperature. The temperature can only drive neutron emission for $S_{\mathrm{n}}\lesssim1.4$ MeV by ($\gamma$,n). Consequently, the density-driven (EC,xn) rates, which are more than a factor of $10^6$ larger than the ($\gamma$,n) rates, expand the path into a band 1.0 MeV $\leq S_{\mathrm{n}} \leq$ 3.5 MeV. Nuclei in this band are generally EC-unstable at  $E_{\mathrm{F}}\gtrsim25$ MeV. Hence, each of the (EC,xn) products (resultant network nodes for different $x$)  are subjected to reaction-driving energies $(E_{\mathrm{F}} - E_{\mathrm{thresh,g.s.}}-S_{\mathrm{xn}})$ that can be higher than in the previous (EC,xn) reaction step, leading to a dense web of subsequent (EC,xn) branchings or a ``Superthreshold (EC,xn) Cascade" (henceforth called the SEC-process) rapidly lowering both $\langle Z \rangle$ and $\langle A \rangle$. For clarity, we denote the absolute value of the g.s.\ nuclear mass differences as $E_{\mathrm{thresh,g.s.}}$ while an actually encountered threshold is always written as $E_{\mathrm{thresh}}$ and includes the energy of the first state to which EC is allowed, which may not be the g.s.

Our network includes EC ($\beta^{+}$) and $\beta^{-}$ rates from
the proton drip-line to the neutron drip-line using transition matrix elements from the
QRPA model described in
\cite{QRPA}. Calculated values of deformations from \cite{mass_def95} were consistently incorporated for $Z>8$ nuclei , below which we have assumed a typical quadrupole deformation $\epsilon_2=0.55$. From any excitation energy $E_{\mathrm{exc}}$ in the EC daughter $(Z-1,A)$  the
neutron decay and competing $\gamma$-ray emission branching ratios
are calculated with the nuclear reaction code GNASH \cite{GNASH}. This code employs the statistical Hauser-Feshbach (HF) method augmented with a Direct Semi-Direct (DSD) component \cite{DSD} for those nuclei in which the direct reactions will dominate. The maximum excitation energy
considered is 30~MeV, and up to 20 emitted neutrons are allowed. We account for neutron kinetic energies which reduce the available energy for subsequent neutron emissions. For $Z<10$  we do not use HF or DSD for the neutron 
branches: rather, if EC occurs into a state above $S_{\mathrm{xn}}$ but below $S_{\mathrm{(x+1)n}}$, then the 
$x$-neutron branch is given full strength and all others are set to zero.




We use radiative neutron capture (n,$\gamma$) rates from \cite{Rauscher}.
The reverse ($\gamma$,n) rate for a typical pre-ND crust nucleus $^{105}$Kr (ignoring partition functions) is $\lambda = (N_{\mathrm{A}} \langle \sigma v \rangle Y_{\mathrm{n}}) \times 10^{10} \times T_9 ^{3/2} \exp(-11.605 S_\mathrm{n}/T_9) = 2.74 \times10^{-6}$s$^{-1}$
for pre-ND neutron abundance $Y_{\mathrm{n}} = 10^{-5}$, a forward (n,$\gamma$) cross-section $N_{\mathrm{A}} \langle \sigma v \rangle =3 \times10^3$ cm$^{3}$g$^{-1}$s$^{-1}$ and $S_\mathrm{n}=1.348$ MeV at crust temperature 0.5 GK. Moving towards a higher $S_\mathrm{n}=2.079$ MeV in $^{101}$Kr  with $N_{\mathrm{A}} \langle \sigma v \rangle = 2\times10^4$cm$^{3}$g$^{-1}$s$^{-1}$, we have $\lambda = 7.82\times10^{-13}$s$^{-1}$ which is already too slow to compete with the accretion timescale $\Delta t \sim10^{9}$s during which the resulting $\Delta E_{\mathrm{F}}$ can make (EC,xn) reactions competitive. 
Typically, only nuclei with  $S_{\mathrm{n}}\lesssim 1.4$ MeV have ($\gamma$,n) competitive with accretion at $T\sim0.5 $ GK.
The products of these 
($\gamma$,n) 
are susceptible to EC which produce $(Z-1)$ nuclei, and the process continues to lower $A$ by further ($\gamma$,n). In models that rely on the stellar photon bath to produce neutrons by ($\gamma$,n) reactions, this will be a valid pathway pre-ND (but constrained by $S_\mathrm{n}\lesssim 1.4$ MeV). In the vicinity of  ND ($E_{\mathrm{F}}\sim25$ MeV), however, the (EC,xn) rates are more than a factor of $10^6$ larger than the ($\gamma$,n) rates. Also, from a heating efficiency perspective, the slow pre-ND ($\gamma$,n) reactions have a negligible effect -- beyond $E_{\mathrm{F}}\sim25$ MeV the SEC-process removes any residual compositional memory. Furthermore, at $E_{\mathrm{F}}\gtrsim20$ MeV the number density of thermal photons is suppressed because the propagating photon is dressed in the medium (the dressed particle is called the plasmon) and acquires a finite mass $\hbar \omega_p = (4 \pi N_e e^2/m_e) (1+(\hbar/(m_e c))^2 (3 \pi^2 N_e)^{2/3} ) \approx $ 1.3 MeV at $\rho Y_e\sim10^{11}$ g/cc increasing to about 4.1 MeV at $10^{12}$ g/cc. Here $Y_e$, $N_e$, $m_e$ are electron fraction, number density and rest mass, respectively. Under typical conditions in the crust ($10^{11}$ g/cc, 0.5 GK) the number density of photons is suppressed by the factor $\exp (-\hbar \omega_p/kT) \simeq 10^{-13}$. Since $\hbar \omega_p \gtrsim S_\mathrm{n}\sim 1.4$ MeV, this suppression has a strong effect at energies important to the pre-ND ($\gamma$,n) reactions. Furthermore, the plasmon suppression increases with increasing density: $\exp (-\hbar \omega_p/kT)  \simeq 10^{-42}$  at $10^{12}$ g/cc and 0.5 GK.
\emph{However, the higher the electron density, the more efficient the (EC,xn) processes become at removing neutrons from increasingly higher excitation energies.}

\emph{Furthermore, the high Nuclear Level Densities (NLDs) at $E_{\mathrm{exc}} \sim S_{\mathrm{xn}}$ and the high superthreshold phase--space factors involved ensure that the SEC-process is a robust mechanism with negligible sensitivity to uncertainties in the underlying nuclear model.} The (EC,xn) timescales will always be shorter than the accretion and the ($\gamma$,n) timescales by several orders of magnitude in a global nuclear model used to calculate the  EC daughter excitation energies. The speed of the resulting reaction flow will be unaffected by the choice of global nuclear model, though the ``transit points'' at a given $E_{\mathrm{F}}$ may differ slightly. We highlight two aspects of the QRPA model we have used which have significant global impact: (1) the single-particle model serving as input to the QRPA is a complete model
space (no levels will be systematically missing) and, (2) we treat deformation realistically. Though global Shell Model calculations of weak interaction strengths for the (heavy and/or neutron--rich) nuclei of interest are simply not available, the same considerations of high NLD at high $E_{\mathrm{exc}} $ apply to nuclear structure computed in the Shell Model as they do to the QRPA calculations.

 \begin{figure}[t]
\includegraphics[width=0.5\textwidth]{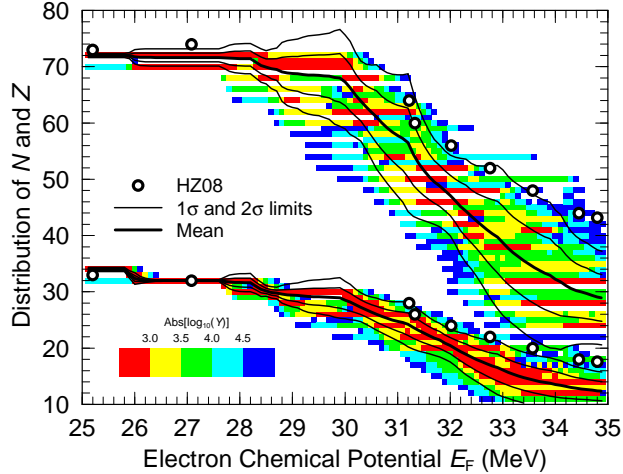}
\caption{(color) Abundance distributions $Y(N)$ and $Y(Z)$ shown for abs$(\log_{10}(Y)) < 5$ at $\Delta E_{\mathrm{F}} = 0.1$ intervals in the $E_{\mathrm{F}}$ range 25--35 MeV for an MCP calculation starting with pure $^{106}$Pd. Beyond $E_{\mathrm{F}}\sim26$ the MCP distributions of $Y(N)$ and $Y(Z)$ rapidly broaden due to EC into highly excited states followed by neutron emissions. The OCP trajectories of \cite{HZ08} starting with the same ICC and evolving through EC and g.s. neutron emissions lie outside the 2$\sigma$ limits of the MCP $Y(N)$ and $Y(Z)$ distributions. }
\label{crust_comp}
\end{figure}

The combined effect of g.s.\ to g.s.\ EC and g.s.\ neutron emissions in the crust was referred to as the ``inverse r-process" by \cite{Sato}. However, the possibility of ``superthreshold" (EC,xn), with very large driving forces that lead to further cascades within a narrow $\Delta E_{\mathrm{F}}$ slice of the crust was not explored. In fact, since earlier models had neutron emission only from g.s.\ they missed important nuclear structure effects in deformed neutron-rich nuclei. The level spacings in these exotic nuclei can influence flows by changing (EC,xn) rates and branchings with only a small change in $E_{\mathrm{F}}$.
Moreover, (1) the SEC-process is quite distinct from the r-process in the Waiting Point Approximation where at a certain temperature $(n,\gamma)-(\gamma,n)$ equilibrium along  constant $Z$ occurs since the (EC,xn) ``reverse rate" is more than a factor of $10^6$ larger than the ($\gamma$,n) rate (even without plasmon suppression); (2) the (EC,xn) rate \emph{increases} with proton-richness, unlike the ($\gamma$,n) rate in the r-process; and (3) the non-thermal reaction energy source $E_{\mathrm{F}}$ is not even approximately constant, as it increases steadily with depth in the NS crust. Therefore, there are few similarities with the r-process, except that that the SEC-process runs in the opposite direction in the $ZN$-plane.

Since the SEC-process covers wide ranges of the nuclear chart with a dense web of competing (EC,xn) channels, it becomes difficult to identify individual nuclei that play a dominant role at a given $E_{\mathrm{F}}$ once the r.m.s. deviation $\Delta N$ increases to $\gtrsim5$ for the abundance distribution $Y(N)$. Rather, we plot the means $\langle N \rangle$ and $\langle Z \rangle$ and the r.m.s. deviations $\Delta Z$, $\Delta N$ (and also $2\Delta N$ and $2\Delta Z$) of the abundance distributions $Y(N)$ and $Y(Z)$ with changing $E_{\mathrm{F}}$ in fig.\ref{crust_comp} to compare our MCP evolution with the OCP evolution of \cite{HZ08} starting with the same initial crust composition (ICC) of pure $^{106}$Pd. We start the evolution at
$E_{\mathrm{F}}\sim0.5$ MeV ($\rho \sim 10^{6}$g/cc) and terminate our simulation when fusion reactions begin to cycle material out of the $Z\sim10$ region at $E_{\mathrm{F}}\sim38$ MeV ($\rho \sim 10^{12.5}$g/cc).

Prior \frenchspacing  to \frenchspacing  neutron \frenchspacing  emission the ``pure EC" path through even-even nuclei is the same as in the OCP calculation of \cite{HZ08}: $^{106}$Pd$\rightarrow^{106}$Ru$\rightarrow^{106}$Mo$\rightarrow^{106}$Zr$\rightarrow^{106}$Sr$\rightarrow^{106}$Kr$\rightarrow^{106}$Se. However, $^{106}$Se$\rightarrow^{106}$As at  $E_{\mathrm{F}}=25.86$ MeV is to $E_{\mathrm{exc}}=2.45$ MeV $ > S_\mathrm{3n}=0.99$ MeV and neutron removal can produce nuclei up to $^{103}$As. Now $E_{\mathrm{thresh}}= 23.46, 24.44, 24.12$ MeV for  $^{103,104,105}$As respectively allow ``superthreshold" (EC,xn) branching with high values of $x$. Thus, $\langle N \rangle$ begins a rapid descent and the MCP trajectory does not produce $N=74$ at  $^{106}$Ge accessed at $E_{\mathrm{F}}\simeq27$ MeV in the OCP (fig.\,\ref{crust_comp}). Well--deformed $^{106}$As ($\epsilon_2 = 0.225$ from \cite{mass_def95}) has neutron energy levels spaced by a few 100 keV, each participating in (EC,xn) over a very small change $\Delta E_{\mathrm{F}}$ in $E_{\mathrm{F}}$. This effect combined with a low $S_{\mathrm{n}}$ results in multiple neutron emission towards (spherical) subshell closures at $N=64, 56$.
Hence, $\Delta N$ suddenly expands to $\sim5$ beyond the subshell $N=70$ and the slope of $\langle N \rangle$ abruptly changes as 
neutrons are rapidly emitted between $N=66$ and 56. We follow the increasingly dense web of (EC,xn) branchings from $^{103}$As(EC,1n)$^{102}$Ge onwards. The product undergoes up to (EC,3n) at $E_{\mathrm{F}}\sim28$ MeV. A dominant branch at 53\% is to $^{101}$Ga whose (EC,xn) products are $^{100-98}$Zn, each with branchings between 30 and 40\%. The flow is now no longer concentrated along a single $x$-neutron emission channel. At $E_{\mathrm{thresh}}=28.6$ MeV $^{98}$Zn(EC,xn) occurs with $x\leq3$. The product $^{95}$Cu ($N=66$) has $E_{\mathrm{thresh}}=28.8$ MeV and releases up to 6 neutrons with substantial branching ratios. Due to this rapidly expanding web of (EC,xn) with increasing $x$, neutrons are emitted continuously as $E_{\mathrm{F}}$ rises, and consequently at $E_{\mathrm{F}}\simeq29$ MeV (fig.\,\ref{crust_comp})  flow from $N=66$ to $N=60$ is already at the factor of 3 level by abundance. In contrast, the OCP essentially does not evolve from $E_{\mathrm{F}}=27.08$ up to 31.22 MeV, i.e. there is no activity between production of $^{106}$Ge and its destruction by $^{106}$Ge+$4e^{-}$$\rightarrow^{92}$Ni+14n+$4\nu_e$ which abruptly releases 14 neutrons at a single $E_{\mathrm{F}}=31.22$ MeV. This scenario is highly unlikely if (1) neutron emissions occur from excited states -- instead of the typical (2EC,6n) steps of the OCP evolution at widely separated $E_{\mathrm{F}}$, (EC,xn) with much higher $x$ can participate in the MCP at several closely spaced $E_{\mathrm{F}}$, and (2) there is more than one network node so that 100\% of the flow is not controlled by a single $x\approx3$ neutron emission mode like a typical OCP (2EC,6n) on $^{80}$Cr at $E_{\mathrm{F}}=32.76$ MeV. If we remove only 3 neutrons after the threshold capture on $^{80}$Cr, the product $^{77}$V is still superthreshold to (EC,xn) by almost 4 MeV. The MCP branchings $\gtrsim 10\%$ on the $^{77}$V intermediate nucleus are 12.6\%(EC,4n), 38.4\%(EC,5n), 15.8\%(EC,6n), 19.2\%(EC,7n) showing odd-even and neutron kinetic energy subtraction effects. The resulting large spreads $\Delta N$, $\Delta Z$ in the MCP evolution allow much smoother movement through several MCP network nodes. This occurs because any one of the network nodes $(Z-1,N+1),(Z-1,N),(Z-1,N-1),...$ resulting from different $x$ in a single (EC,xn) step can itself move on to lower $N$ and $Z$ through (EC,xn) either at the prevailing $E_{\mathrm{F}}$ or with a small increase $\Delta E_{\mathrm{F}}\sim 0.3$ MeV.


The nuclear structure probed by the high energy $(E_{\mathrm{F}} - E_{\mathrm{thresh}})$ of ``superthreshold (EC,xn)" determines the characteristics of the reaction flow in the following ways:

(1)\emph{The slopes of $\langle N \rangle , \langle Z \rangle$ with $E_{\mathrm{F}}$, or the speed of the reaction flow, comes from a removal of the (spherical) subshell degeneracy due to deformation. This effect is strongest midway between spherical (sub)shells}. The OCP curve passes through $N= 64, 60, 56, 52, 48, 44, 40, 36, 30$ with large gaps $\Delta E_{\mathrm{F}}\sim1$ MeV(and a huge gap of 4.14 MeV between the productions of $^{106}$Ge and $^{92}$Ni). In contrast, the MCP neutron emission is much more continuous with changing $E_{\mathrm{F}}$.
At higher $E_{\mathrm{F}}$ in the MCP evolution, fig.\,\ref{crust_comp} clearly shows that within the (spherical) neutron $g_{9/2}$ subshell ($N$ between 40 and 50) there are several transit points in $N$
 that become accessible at very closely spaced $E_{\mathrm{F}}$. \emph{These correspond to the well-separated deformed neutron levels emanating from the same (spherical) subshell}. A level diagram of mid-proton-shell $^{80}$Cr (deformed with $\epsilon_2=0.233$) not only shows this effect, but also that upon entry into the $N=56$ subshell closure, $N=54,52$ should play a prominent role and  $N=52,50$ correspond to levels at almost identical energy. Therefore, the MCP $N$-evolution flattens at $N=56$ and 40, but not at $N=50$, through which it is rather steep. 
Throughout the evolution the OCP slope of $N(E_{\mathrm{F}})$ (with notable exceptions at $N=64,60$ corresponding to $Z=28,26$) is closer to the MCP slopes of $\langle N \rangle (\pm 2\Delta N)$ \emph{between} major neutron (sub)shells (see the MCP slopes at $N=56,40$ in fig.\,\ref{crust_comp}) -- thus the OCP g.\frenchspacing s. \frenchspacing neutron removals do not reflect the rich structure within neutron (sub)shells for well-deformed neutron-rich nuclei.

(2) \emph{$\Delta N$ and $\Delta Z$ increase to reflect the occupancies of accessible (sub)shells at a given $E_{\mathrm{F}}$}. An example is the increase of $\Delta N$ to $(40-28)/2=6$ closely matching the network value of 6.44 at $E_{\mathrm{F}}=33.5$ MeV.  Levels beneath subshell closures at $N=38,32$ are all accessible at this $E_{\mathrm{F}}$ but $N$ between 28 and 20 is only accessed at $N=26$ by highly abundant nuclei.

(3)\emph{Unlike the OCP evolution, branchings populate odd-$A$ chains also -- these are rapidly depleted, driving
abundance to lower $Z$ faster}. 

(4)\emph{The free neutron abundance $X_{\mathrm{n}}$ increases smoothly with $E_{\mathrm{F}}$ due to the large number of available (EC,xn) sources at each depth in marked contrast to the discontinuous evolution of $X_{\mathrm{n}} (E_{\mathrm{F}})$ in the OCP (see \cite{HZ90,HZ03,HZ08}).} 


The resulting distribution $Y(N)$ at any $E_{\mathrm{F}}$ is multiple-peaked primarily due to odd-even and neutron (sub)shell effects. It is also highly asymmetric with long neutron-deficient tails that are susceptible to further (EC,xn).  \emph{This leads to a broadening of $Y(Z)$ to the extent allowed by the accessible proton (sub)shells.}

Beyond $E_{\mathrm{F}}\sim29$ MeV the MCP rapidly diverges from the OCP since roughly $[(2 \Delta Z)\times (2 \Delta N)] \approx 5\times 10 = 50$ network nodes are simultaneously processing the abundances unlike a single node in an OCP scenario.  This dense Markov process of (EC,xn) transitions (1) quickly removes compositional memory of the XRB ashes through several $N$-nodes at each $Z$ (and vice-versa), which is not possible in an OCP evolution, and (2) changes the evolutionary description from one of individual nuclei at specific $E_{\mathrm{F}}$ (which is possible for the OCP) into that of r.m.s. deviations and higher moments which ultimately determine the shapes of the $Y(N)$ and $Y(Z)$ distributions. These shapes are clearly discernible in fig.\,\ref{crust_comp} since the color changes denote abundance variations by a factor of only $10^{0.5}\simeq3$.
The large spreads $\Delta N$ and $\Delta Z$ in these distributions are important inputs to models of transport processes in the crust such as those determining thermal conductivity.

\begin{figure}[t]
\includegraphics[width=0.5\textwidth]{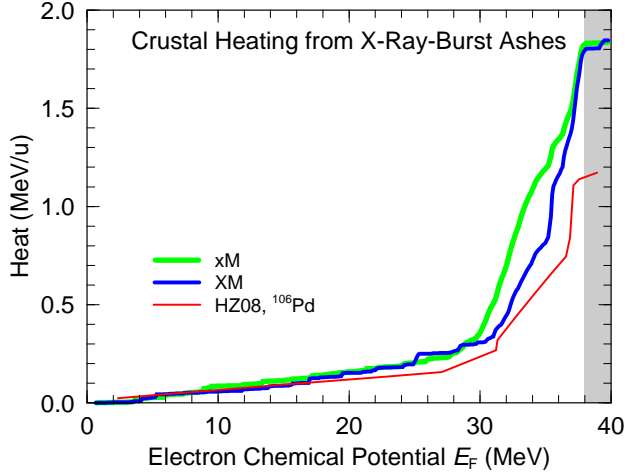}
\caption{(color) Crustal heating in our MCP model for ICC (XRB ashes from \cite{WH04}) from NS accreting 5 \% solar ($x$) and solar metallicities ($X$) at the accretion rate $\mathrm{M} = 0.1 \dot{m}_{\mathrm{Edd}}$ characteristic of Superburst progenitors (accretion at the Eddington limit is $\dot{m}_{\mathrm{Edd}}\approx8.85\cdot10^4$gcm$^{-2}$s$^{-1}$).  Also plotted are OCP heating from \cite{HZ08} for the ICC comprised of $^{106}$Pd. 
\hspace{1.0in}\emph{ICC from multizone XRB model xM}: Peak $A\sim100-110$ (very similar to the 1-zone rp-process
products of \cite{schatz} and also to an ICC of pure $^{106}$Pd in the OCP model of \cite{HZ08}). 
\hspace{1.0in}\emph{ICC from model XM}: A bimodal abundance distribution with significant pre-Fe-peak abundance, and also in $A\sim60-80$.}
\label{crust_heat}
\end{figure}



\printfigures

The efficiency of the SEC-process in heating the crust is shown in fig.\,\ref{crust_heat}. We have the remarkable result that MCP crustal heating profiles from two very different ICC from possible Superburst progenitors change slope drastically near  $E_{\mathrm{F}}\sim35$ MeV and converge to within $\sim0.1$ MeV/u of each other. This is of the order of the difference in BE/A of the ICC. Note that the MCP heating diverges from the OCP at a depth $E_{\mathrm{F}}\simeq30$ MeV where the MCP slopes of $\langle N \rangle (\pm 2\Delta N)$ change dramatically in fig.\,\ref{crust_comp}.

The bulk of deep crustal heating by the SEC-process occurs in a narrow region post-ND, rather than near the crust-core interface if pycnonuclear fusion dominates. Being further from the core this new crustal heating mechanism can result in thermal profiles conducive to the $^{12}$C ignition of X-ray Superbursts. Furthermore, unless the ICC is comprised only of pre-Fe-peak nuclei concentrated in $A=36-44$ (fig.\,7 of \cite{Gupta} explains why shallow heating should differ in this case; however the total crust heating will still differ by $\lesssim0.2$ MeV/u from MCP evolution of other ICC), the composition as a function of depth is not dependent on the ICC. Thus for a large class of accreting NS, the composition-dependent conductivities and crust neutrino cooling will look very similar, and so will the crustal thermal profiles which depend on these quantities.

We thank S. Reddy for helpful discussions and for suggestions for improvement after a careful reading of the manuscript.
We thank A. Heger for providing the X-ray burst ashes of \cite{WH04}
 for use in our calculations. We are grateful for the encouraging and helpful
advice from the referees.
This work was funded in part by the LDRD program at LANL under grant number 20080130DR.


\bibliography{crust_prl}
\end{document}